\documentclass[aps,prl,twocolumn,tightenlines,showpacs,nofootinbib]{revtex4}
\usepackage{bm}
\usepackage{graphics}
\usepackage{epsfig}
\usepackage{graphicx}
\usepackage{amsmath}
\usepackage{amsfonts}
\usepackage{amssymb}
\begin{document}

\title{Hysteresis loop signatures of phase transitions in a mean-field model of disordered Ising magnet}
\author{P. N. Timonin}
\email{timonin@aaanet.ru}
\affiliation{Physics Research Institute at
Southern Federal University, 344090 Rostov-on-Don, Russia}

\date{\today}

\begin{abstract}
In accordance with recent experiments the mean-field type theories predict the presence of numerous metastable minima (states) in the rugged free-energy landscape of frustrated disordered magnets. This multiplicity of long-lived states with lifetimes greater than $10^5 s$ makes the task to experimentally determine which of them has the lowest free energy (and thus what thermodynamic phase the sample is in) seem rather hopeless the more so as we do not know a protocol (such as field-cooling or zero-field-cooling) leading to the equilibrium state(s). Nevertheless here we show in the framework of Landau-type  phenomenological model that signatures of the mean-field equilibrium phase transitions in such highly  nonequilibrium systems may be found in the evolution of the hysteresis loop form. Thus the sequence of transitions from spin-glass to mixed phase and to ferromagnetic one results in the changes from inclined hysteresis loop to that with the developing vertical sides and to one with the perfectly vertical sides. Such relation between loop form and the location of global minimum may hold beyond the mean-field approximation and can be useful in the real experiments and Monte-Carlo simulations of the problems involving rugged potential landscape. Also the very existence of the quasi-static loops in spin glass and mixed phases implies that the known disorder-smoothing of the first-order transition can be always accompanied by the emergence of multiple metastable states.     
\end{abstract}


\maketitle

Many types of disorder can induce the frustration in the magnetic interactions of a solid. This results in appearance of the glassy states in disordered magnets which may exist along with the more usual ones - ferromagnetic, ferrimagnetic etc. There are also mixed phases where perfect magnetic order is destroyed only partially. This variety of thermodynamic phases (states) is present in many microscopic models with frozen frustrated disorder \cite{1, 2, 3, 4, 5}. Yet an equilibrium thermodynamic phase may be something like a phantom in such highly nonequilibrium systems. From Monte-Carlo studies of microscopic spin-glass models \cite{6} it is known what a huge time is needed for such systems to relax into the equilibrium state. From the mean-field point of view this is the consequence of the rugged free energy landscape with numerous local metastable minima in each of which the system can be trapped on its way to the global (equilibrium) minimum for a time which grow exponentially with the volume of a sample \cite{7}. Recent experiments \cite{8, 9} show visually that in zero and finite field random magnets do stay confined in the variety of the phase space regions with different magnetizations for times up to $10^5 s$ without a noticeable relaxation. 

Moreover, Monte-Carlo simulations of short-range 3d Ising spin-glass under quasi-static field variations \cite{10} demonstrate the existence of numerous stationary states inside the hysteresis loop. These findings are shown to be in a reasonable qualitative agreement with the similar experiments in the Ising spin-glass $Fe_{0.5}Mn_{0.5}TiO_3$ \cite{10}. Along with the experiments of Refs.\cite{8, 9} this is the unambiguous evidence that short-range Ising spin-glasses do can posses the multi-valley structure of phase space even in external field. 
  
Often the validity of such mean-field picture for the short-range spin-glasses is questioned on the grounds of experiments (Ref. \cite{11} and references therein) and simulations \cite{12, 13} which do not find any Almeida-Thouless anomalies in a field. It seems that the imaginary paradox between evidences of the multi-valley structure \cite{8, 9, 10} and the absence of Almeida-Thouless anomalies \cite{11, 12, 13} stems from the unjustified conviction that in short-range systems the onset of spin-glass phase should be necessarily accompanied by the divergence of the spin-glass susceptibility same as in the Sherrington-Kirkpatrick model.

So the rather natural resolution of this paradox lies in the recognition that in the short-range spin-glasses the multi-valley landscape can emerge in a finite field without the Almeida-Thouless anomalies. The qualitative explanation of such behavior can be found in the framework of multi-mode condensation mechanism \cite{14, 15, 16} describing the possible origin of the spin-glass phase in short-range random Ising magnets. Indeed, in this mechanism the whole macroscopic set of sparse fractal modes become unstable at the same transition temperature in zero field while in finite field their instability points spread throughout a finite temperature interval, cf. Fig. 4 in Ref. \cite{16}. So the experiments and simulations in the field would not detect the sharp transition, instead the gradual emergence of more and more corrugated landscape would take place in the larger temperature intervals for higher fields. 

Thus we may conclude that the mean-field viewpoint is applicable to the real random Ising magnets at least on the present days' laboratory times. The last reservation is important as usually the mean-field theories overestimate the barriers between local minima and, hence, the Arrhenius lifetimes of the metastable states. Thus the Monte-Carlo simulations of short-range 3d Ising spin-glass \cite{17, 18} indicate that there can be the directions in phase space where the barrier's heights stay finite. Hence the mean-field notion of the divergent metastable states' lifetimes can be violated at longer time scales, that is the system can relax to the global minimum after, say, months or years. 

Yet, being in error with respect to the barriers and lifetimes, some refined mean-field models may still give correct relative depths of metastable minima in the free-energy landscape. In such cases we would have a correct mean-field description of a system's true phase diagram which would agree with the exact results of the equilibrium statistical mechanics valid at infinite times. Thus the mean-field approach may be useful in the locating the points of the true equilibrium transitions.

So with the above reservations we may interpret the experimental data in the mean-field regime spreading up to the times of order $10^5 s$ with the notion of metastable states with extremely long and divergent lifetimes. Hence, we are faced with the problem to find among the variety of the observed stationary states the equilibrium one(s) whose properties are described by the equilibrium microscopic models. In the absence of some notion of a possible protocol (such as field-cooling or zero-field-cooling) leading to the equilibrium state(s) it is rather hopeless task if one is not prepared to wait for months and years in (probably vain) hope that the mean-field predictions fail at longer times and our random system still find the way to the equilibrium state. 

Nevertheless the definite signatures of equilibrium thermodynamic transitions in some glassy magnets may be found in quite simple experiments. They can be encoded in the temperature evolution of their hysteresis loop form \cite{15, 19}. This is quite natural as the presence of hysteresis loop is the main and unambiguous manifestation of the irreversibility appearing in glassy phases. Due to the mean-field picture the origin of this irreversibility lies in the presence of multiple metastable states so one can easily conclude that their magnetization curves should lie in the loop interior. This apparent conclusion needs no further theoretical justification once the existence of metastable states is adopted. Hence this should be the necessary feature of any adequate mean-field model of glassy magnets.

In spite of its simplicity the notion of the loop filled with the metastable magnetization curves allows to explain qualitatively the origin of all quasi-static irreversible phenomena in glassy magnets. Thus assuming that in the spin-glass phase the observed inclined hysteresis loops are filled with the metastable magnetization curves one can easily explain how the different values of physical quantities emerge in various experimental protocols \cite{16}. For example, it becomes quite obvious that the field-cooled process brings the system to the upper outline of the loop while the zero-field-cooled one makes it join its lower outline \cite{16}.  

Moreover the inspection of hysteresis loop form can give some notion on the relative depths of the metastable minima thus making a link with the predictions of the equilibrium statistical mechanics. On this way the quite plausible assumption is that the deeper zero-field metastable minimum the larger coercive field of its magnetization curve (at which the minimum vanishes destroyed by the field opposite to the zero-field magnetization). As the ends of the metastable magnetization curves constitute the outline of the major hysteresis loop one can deduce from its form which of the metastable state has the lowest zero-field free energy thus being the equilibrium one. Such heuristic considerations for the determination of phase sequence in glassy systems were first advanced in Ref.\cite{19}. Here we show that in the framework of the Landau-type phenomenological model the equilibrium phase transitions do show up in the hysteresis loop form and this heuristics is exact in spin-glass and ferromagnetic phases and approximate in the mixed one.
	
We consider the Landau potential for the magnetizations of the sparse fractal modes $m_i$, $i=1$,...,$N_0$ \cite{9}, \cite{10} of the form
\begin{eqnarray*}
F\left( {\bf{m}} \right) = \frac{{\tau _g }}{2}\left( {\left[ {m^2 } \right] - \left[ m \right]^2 } \right) + \frac{{\tau _f }}{2}\left[ m \right]^2  + \frac{1}{4}\left[ {m^4 } \right]
\\
 + \frac{d}{4}\left[ m \right]^4  - h\left[ m \right]
\end{eqnarray*}
Here $\left[ {m^k } \right] = N_0^{ - 1} \sum\limits_{i = 1}^{N_0 } {m_i^k }$. 
It is the simplest model suitable for our aim as at $d>0$ it has three magnetic phases (spin-glass, mixed and ferromagnetic) possible for the Ising-type ferromagnet with frustrated nonmagnetic disorder. In the multiple local minima of $F\left( {\bf m} \right)$ $m_i$ acquire just two values $m_+>0$ and $m_-<0$ so all their variety is parameterized by the quasi-continuous variable $0<n<1$ defining the fraction of the positive $m_i$ at a given metastable minimum. The existence of these minima is limited by condition $\tau _g  < 0$ and the stability conditions
\begin{equation*}
m_ +  \left( n \right)/2 <  - m_ -  \left( n \right) < 2m_ +  \left( n \right), {\rm   }
\chi \left( n \right) \equiv \frac{{\partial m\left( n \right)}}{{\partial h}} > 0
\end{equation*}
Here 
\begin{equation}
m\left( n \right) = nm_ +  \left( n \right) + \left( {1 - n} \right)m_ -  \left( n \right)
\label{eq:1} 
\end{equation} 
is the net magnetization of a sample in the metastable state and $\chi \left( n \right)$ is its magnetic susceptibility.
Among these local minima those with the lowest free-energy (equilibrium states) are defined by the conditions
\begin{eqnarray}
m_ +  \left( n \right) + m_ -  \left( n \right) = 0, \label{eq:2}
\\
\tau _f  + \max {\left( - \tau _g , m\left( n \right)^2 \right)}  + 3dm\left( n \right)^2  > 0, \qquad  m\left( n \right)h > 0. \nonumber
\end{eqnarray}

 The magnetization of equilibrium states $m_{eq}$ obeys the equation
\begin{equation}
\left[ \tau _f  + \max \left( { - \tau _g ,m_{eq}^2} \right) \right]m_{eq}  + dm_{eq}^3  = h
\label{eq:3}
\end{equation}
and the parameter $n=n_{eq}$ of these states is
\begin{equation}
n_{eq}  = \frac{1}{2}\left[ 1 + sign\left({ m_{eq}} \right)\sqrt{ \min \left( {\frac{m_{eq}^2 }{ - \tau _g },1} \right)} \right]. 
\label{eq:4}
\end{equation}
Also for the equilibrium value of spin-glass Edwards-Anderson order parameter
\[
q\left( n \right) = \left[ {m^2 } \right] - \left[ m \right]^2  = n\left( {1 - n} \right)\left[ {m_ +  \left( n \right) - m_ -  \left( n \right)} \right]^2 
\]
we get from Eqs.(\ref{eq:1}),(\ref{eq:2}), (\ref{eq:4})
\[
q_{eq}  =  \min (- \tau _g  - m_{eq}^2, 0) 
\]
In zero field at $\tau _g  < \tau _f$, $\tau _g  < 0$ we have 
\[
m_{eq}  = 0,\qquad n_{eq}  = 1/2, \qquad q_{eq}  =  - \tau _g 
\]
so here the spin-glass (SG) phase is realized.  At $h \to  + 0$ and $\left( {1 + d} \right)\tau _g  < \tau _f  < \tau _g  < 0$ the mixed (M) phase with partially ordered moments exists with
\begin{eqnarray}
m_{eq}  = \sqrt {\frac{{\tau _g  - \tau _f }}{d}} ,\qquad 
n_{eq}  = \frac{1}{2}\left( {1 + \sqrt {\frac{{\tau _f  - \tau _g }}{{d\tau _g }}} } \right), \label{eq:5}
 \\
q_{eq}  = \left[ {\tau _f  - \tau _g \left( {1 + d} \right)} \right]/d
 \nonumber
\end{eqnarray}
At last for $h \to  + 0$ and $\tau _f  < \left( {1 + d} \right)\tau _g  < 0$
 we have ferromagnetic (FM) phase where
\[
m_{eq}  = \sqrt {\frac{{ - \tau _f }}{{1 + d}}},\qquad
n_{eq}  = 1,\qquad q_{eq}  = 0.
\]
In spite of continuous variation of $m_{eq}$, $n_{eq}$ and $q_{eq} $
throughout all phases the transitions between them are of the first-order. This is because the SG and FM phase stay always stable at SG-M and M-FM transition points correspondingly.

\begin{figure}[htp]
\centering
\includegraphics[height=6cm]{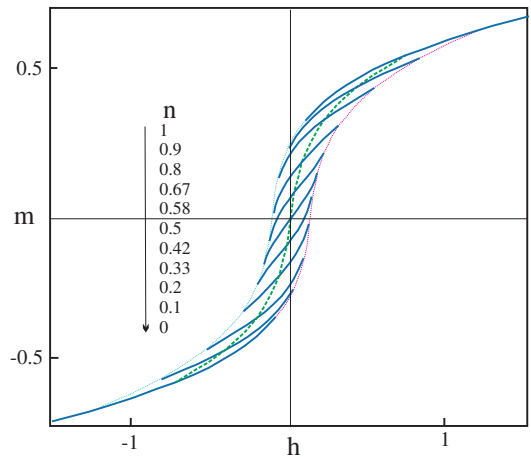}
\caption{(color online) Hysteresis loop formed by metastable magnetization curves in SG phase at $\tau_g=-0.45$. Dotted lines show the stability limits of metastable states, dashed line indicates the lowest (equilibrium) minima.  }\label{Fig.1}
\end{figure}

\begin{figure}[htp]
\centering
\includegraphics[height=5.5cm]{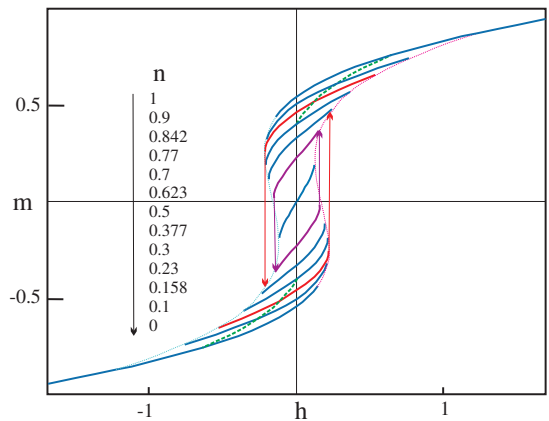}
\caption{(color online) The same as in Fig.1 for M phase at $\tau_g=-0.58$.}
\label{Fig.2}
\end{figure}

\begin{figure}[htp]
\centering
\includegraphics[height=5.5cm]{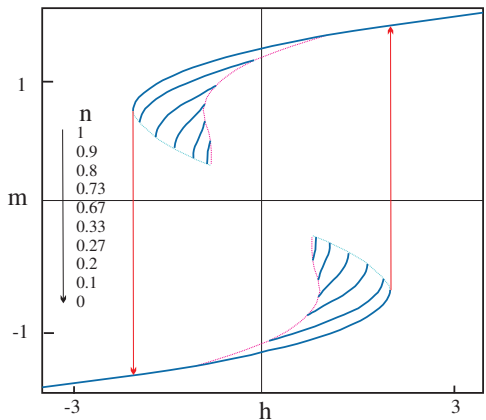}
\caption{(color online) Hysteresis loop in F phase at $\tau_g=-1.25$.} \label{Fig.3}
\end{figure}

Now we turn to the temperature evolution of the hysteresis loop accompanying these transitions. As both $\tau _g$ and $\tau _f$ are linear functions of temperature in a specific sample there is a linear relation between them $\tau _f  = c\tau _g  + \tau _0$ with $c,{\rm{ }}\tau _0  = const$. Choosing $c>d+1$ and $\tau _0  > 0$ we get the system with the SG-M-FM phase sequence when $\tau _g $ diminishes. Figs.\ref{Fig.1}-\ref{Fig.3} show the subsequent evolution of the hysteresis loop filled with metastable curves through such phase sequence. Here $c=5$, $\tau _0  = 2$ $d=2$ so the transition temperatures correspond to $\tau _g  =  - \tau _0/(c - 1) =  - 0.5$ (SG-M) and $\tau _g  =  -\tau _0/(c - d - 1) =  - 1$ (M-FM). 
 
 In SG phase we have inclined loop ubiquitous in random magnets (Fig.\ref{Fig.1}) while in M phase the region of stability of metastable states acquires the S-shaped form which results in the development of vertical jumps in the major loop (Fig.\ref{Fig.2}). This takes place right at SG-M equilibrium transition point and the critical fields $ \pm h_s $ and magnetizations $ \pm m_s $ at which the jumps start are defined through

\begin{equation*}
m_s  = \sqrt {\frac{{\tau _g  - \tau _f }}{{3d }}}, 
\qquad 
h_s  = 2\left[ {\left( {\frac{{ - \tau _g }}{3}} \right)^{3/2}  + d m_s^3 } \right].
\end{equation*}

The magnetizations $\pm m_s$ are strictly proportional to the zero field equilibrium magnetization $m_{eq}$ (cf. Eq. (\ref{eq:5}).
%xxxxxxxxxxxxxxxxxxxxxxxxxxxxxxxxxxxxxxxxxxxxxxxxxxxxxxxxxxxxxxxxxxxxxxxxxxxxxx
%xxxxxxxxxxxxxxxxxxxxxxxxxxxxxxxxxxxxxxxxxxxxxxxxxxxxxxxxxxxxxxxxxxxxxxxxxxxxxx. 
%xxxxxxxxxxxxxxxxxxxxxxxxxxxxxxxxxxxxxxxxxxxxxxxxxxxxxxxxxxxxxxxxxxxxxxxxxxxxxx
%xxxxxxxxxxxxxxxxxxxxxxxxxxxxxxxxxxxxxxxxxxxxxxxxxxxxxxxxxxxxxxxxxxxxxxxxxxxxxx. 
%xxxxxxxxxxxxxxxxxxxxxxxxxxxxxxxxxxxxxxxxxxxxxxxxxxxxxxxxxxxxxxxxxxxxxxxxxxxxxx
%xxxxxxxxxxxxxxxxxxxxxxxxxxxxxxxxxxxxxxxxxxxxxxxxxxxxxxxxxxxxxxxxxxxxxxxxxxxxxx. 
Such transitions in the loop form are found in experiments \cite{19, 20, 21} and in various random spin models \cite{22, 23, 24}. We may reasonably suppose that they are the manifestations of the equilibrium SG-M transition in them.

We can also mention another specific feature of M phase, that is the minor loops with vertical sides which tops (bottoms) are composed of metastable curves with sufficiently large $n$ ($1-n$), see Fig.\ref{Fig.2}. At the M-FM transition point $m_s $ reaches the stability point of the homogeneous state ($n=1$) so the loop in F phase becomes the ordinary ferromagnetic one composed entirely of homogeneous magnetic states (Fig.\ref{Fig.3}). The metastable curves inside it are inaccessible in quasi-static regime yet they can be reached with special fast field changes. 

The dashed lines in Figs. \ref{Fig.1}, \ref{Fig.2} show the position of the deepest equilibrium minima at a given field thus comprising the results for the equilibrium magnetization curves which can be obtained in the (mean-field) models of the equilibrium statistical mechanics. Apparently they are unobservable in the real-time experiments. But if one manage to get the experimental data in ac field with period much greater than $10^5 s$ the shrinking of loops to the equilibrium curves in Figs. \ref{Fig.1}, \ref{Fig.2} may be observed. If this will be actually the case and at what superlong times this may occur is still the open question of modern theory. In terms of the equilibrium curves the main feature of the present model can be characterized as simultaneous qualitative changes of these curves and the loop's outline, e. g. the jumps at the loop's sides appear simultaneously with the jump in the equilibrium curve at $h=0$ indicating the onset of the mixed phase. The real magnets and more realistic models may not have such strict correspondence yet on the basis of above heuristics we may expect in them qualitatively similar behavior. We may also note that in random anisotropic Heisenberg magnets the evolution of loops would depend on the field direction and the variety of mixed and spin-glass phases can be detected from this dependence.  

To elucidate the physical content of the present phenomenology we note that it still holds for the ordinary ferromagnets with low concentration of impurities and defects. Here the metastable states are just the numerous multi-domain patterns stable in some intervals of applied field. Each pattern has its own magnetization curve along which the net magnetization varies due to the change of the domain's moments. At the ends of this curve the pattern looses its stability and transforms into another one via domain wall jumps. The evolution of the hysteresis loop filled with these curves in such slightly disordered ferromagnet will follow that depicted in Figs.\ref{Fig.1}-\ref{Fig.3} as with lowering temperature the gradual growth the domain wall energy will make lower the potential of the states with fewer walls and larger net moment. With increasing disorder the domains in these patterns will be smeared but nevertheless their spin configurations still will be stable corresponding to the local free energy minima. Thus the metastable spin-glass states can be viewed as smeared domain patterns. In the absence of domain walls the transitions between them are realized through the avalanches of spin upturns on the sparse fractal set of sites belonging to the specific eigenmode of random exchange matrix \cite{15, 16}. Hence Barkhausen noise and acoustic emission can be also observed at the sides of the outer loops in glassy phases.

 Figs.\ref{Fig.1}-\ref{Fig.3} shed also a new light on the smearing of the first order transitions by a quenched disorder \cite{25}. Indeed, in perfect ferromagnet the field sweeping from negative to positive values results in the first order transition in which magnetization direction is reversed. So Figs. \ref{Fig.1}, \ref{Fig.2} show what smearing effect the disorder can have on this transition. Here the important feature consists in the appearance of multiple metastable states in the vicinity of smoothed transition. It is quite probable that this is inherent property of all smeared first order transitions. Thus multiple magnetization curves show up near the AFM-FM transition in Fe-Rh alloy \cite{26}. 

\begin{figure}[htp]
\centering
\includegraphics[height=4.5cm]{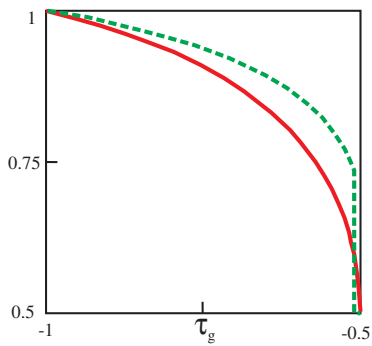}
\caption{(color online) Temperature dependence of $n_{eq}$ (solid line) and $n_c$ (dashed line) in M phase}\label{Fig.4}
\end{figure}

Here we can verify do the deepest (equilibrium) zero-field state with $m_{eq} \ge 0$ ($n_{eq}  \ge 1/2$) has the largest module of its negative coercive field, i.e. the field destroying the local minimum. From  Figs. \ref{Fig.1}, \ref{Fig.3} we see that it is evidently true in SG and F phases. Designating the parameter $n$ of the state with the largest module of negative coercive field as $n_c$ we have $n_c  = n_{eq}  = 1/2$ in SG phase and $n_c  = n_{eq}  = 1$ in F phase. In M phase the $n_c$ of the state with largest coercive field is 
\[
n_c  = \left\{ \begin{array}{l}
 1/2,{\rm{ }}\tau _g \left( {1 + 0.037d} \right){\rm{ <  }}\tau _f  < \tau _g  \\ 
 \frac{1}{3}\left( {2 + \sqrt {\frac{{\tau _f  - \tau _g }}{{d\tau _g }}} } \right),{\rm{ }}\tau _g \left( {1 + d} \right){\rm{ < }}\tau _f {\rm{ <  }}\tau _g \left( {1 + 0.037d} \right) \\ 
 \end{array} \right.
\]
This $n_c$ does not coincide with $n_{eq}$ in Eq. (\ref{eq:5}) yet their values stay rather close throughout M phase as Fig. \ref{Fig.4} shows. So here the above heuristics works only approximately. 

Nevertheless the peculiarities of the major hysteresis loop form in the present model strictly follow the equilibrium transitions. The thermodynamic phases in such system can be identified through simple experiments in the quasi-static periodic field varying sufficiently slowly to follow the potential profile. It can take much less time than the relaxation into the deepest minimum as encircling major loop the system sweeps the boundary between the rough and smooth landscapes and the relaxation times here would  not be that enormous. Yet the strict relation between loop form and equilibrium transitions can be lost in more realistic models with complicated potential landscapes. One may argue that possible deviations from the above picture should not be large as present model gives quite adequate description of the loop evolution in real glassy objects \cite{19, 20, 21} and the underling heuristic considerations can still be valid for the real systems. 
%xxxxxxxxxxxxxxxxxxxxxxxxxxxxxxxxxxxxxxxxxxxxxxxxxxxxxxxxxxxxxxxxxxxxxxxxxxxxxx
%xxxxxxxxxxxxxxxxxxxxxxxxxxxxxxxxxxxxxxxxxxxxxxxxxxxxxxxxxxxxxxxxxxxxxxxxxxxxxx
Thus it is worthwhile to study with Monte-Carlo simulations the relations between the loop form and the global minimum position in the microscopic short-range models. Moreover just these studies can save the results of the equilibrium statistical mechanics and Mont-Carlo absolute minima searches from being useless for the real experiments. Figs.\ref{Fig.1}-\ref{Fig.3} show unambiguously that full knowledge of the potential landscape and its dependence on the field and temperature are necessary for understanding the quasi-static processes in glassy systems. It would not be a formidable task for experiments and simulations as it just comes to the determination of the temperature evolution of hysteresis loop and metastable curves inside it. In reward one gets the ability to describe every (unavoidably nonequilibrium) process on the experimental time scales \cite{15, 16}.

Also this strategy and heuristic relation between the coercive fields and minima depths can be used in the variety of optimization problems once a suitable couple of conjugate variables playing the role of $m$ and $h$ can be found.  
For example, it is quite evident from Figs. \ref{Fig.1}, \ref{Fig.2} that the ground state search with numerical version of demagnetization process \cite{27} can converge to the non-magnetized ground state but should be modified for the magnetized one. So the sensible ground state search should include preliminary determination of the hysteresis loop form as have been already noted in Ref. \cite{23}. 

%------------------------------------------------------------------------------

%
%
\end{document}